\documentclass[preprint,12pt,authoryear]{elsarticle}

\usepackage{graphicx}
\usepackage{amssymb}
\usepackage{amsmath}
\usepackage[colorlinks=true, linkcolor=blue, citecolor=red, urlcolor=blue]{hyperref}
\usepackage{natbib}
\bibpunct{(}{)}{;}{a}{}{,}

\journal{New Astronomy Reviews}

\begin{document}

\begin{frontmatter}



\title{A new view on the ISM of galaxies: far-infrared and
  submillimetre spectroscopy with Herschel}


\author[UGent]{Maarten~Baes}
\author[Paris]{Suzanne~C.~Madden}
\author[McMaster]{Christine~D.~Wilson}
\author[ESOChile]{Andreas~A.~Lundgren}
\author[ESOGermany]{Carlos~De~Breuck}

\address[UGent]{Sterrenkundig Observatorium, Universiteit Gent,
%  Krijgslaan 281 S9, B-9000 Gent, 
Belgium}
\address[Paris]{Service d'Astrophysiques, CEA/Saclay, 
%L'Orme des Merisiers, 91191 Gif-sur-Yvette, 
France}
\address[McMaster]{Department of Physics and Astronomy, McMaster University, %Hamilton, ON L8S 4M1, 
Canada}
\address[ESOChile]{European Southern Observatory, 
%Casilla 19001, 
Santiago, Chile} 
\address[ESOGermany]{European Southern Observatory, 
%Karl-Schwarzschild Strasse, 
Garching bei M\"unchen, Germany}

% \author{Maarten~Baes} 
% \address{Sterrenkundig Observatorium, Universiteit Gent, Belgium}
% \author{Suzanne~C.~Madden}
% \address{Service d'Astrophysiques, CEA/Saclay, France}
% \author{Christine~D.~Wilson}
% \address{Department of Physics and Astronomy, McMaster University, Canada}
% \author{Andreas~A.~Lundgren}
% \address{European Southern Observatory, Santiago, Chile} 
% \author{Carlos~De~Breuck}
% \address{European Southern Observatory, Garching bei M\"unchen, Germany}

\begin{abstract}
  The FIR/submm window is amongst the least explored spectral regions
  of the electromagnetic spectrum. It is, however, a key to study the
  general properties of the interstellar medium of galaxies, as it
  contains important spectral line diagnostics from the neutral,
  ionized and molecular ISM. The Herschel Space Observatory,
  successfully launched on 14 May 2009, is the first observatory to
  cover the entire FIR/submm range between 57 and 672~$\mu$m. We
  discuss the main results from the ISO era on FIR spectroscopy of
  galaxies and the enormous science potential of the Herschel mission
  through a presentation of its spectroscopic extragalactic key
  programs.
\end{abstract}


\end{frontmatter}


\section{Introduction: the ISM of galaxies}

The ISM of galaxies is not a quiescent medium; it lives in a constant
mutual interaction with the stellar component of a galaxy. On the one
hand, the ISM is the birthplace of the stars; on the other hand, the
stars control the structure and therefore the star formation rate of
the ISM. In general, the ISM consists of three distinct phases
(components that live in relative thermal pressure equilibrium): a
cold, a warm and a hot phase. Dotted between these phases are dense,
mainly molecular cores, either collapsing to form stars or expanding
due to the feedback of embedded or nearby stars. Most of the mass of
the ISM is in neutral regions, either in the cold neutral gas or the
molecular clouds. With the exception of the molecular gas deep in the
cores of dense star-forming clouds, the physics, chemistry and
evolution in these neutral regions is dominated by stellar
far-ultraviolet (FUV) photons, with energies between 6 and 13.6
eV. Such regions are generally called photo-dissociation regions
(PDRs).  Detailed review papers on the different aspects of PDRs can
be found in e.g.\ \citet{1997ARA&A..35..179H, 1999RvMP...71..173H}.

Not only do PDRs include most of the mass of the ISM in galaxies, they
also are responsible for the bulk of the FIR/submm radiation of
galaxies. The incident FUV starlight on PDRs is absorbed primarily by
dust grains and PAHs. The vast majority of this absorbed energy is
re-emitted as FIR/submm continuum radiation or PAH line emission. A
minor fraction of the energy, typically 0.1 to 1\%, is converted to
energetic photoelectrons, which heat the PDR gas, a process known as
photoelectric heating. The gas is less efficient at cooling than the
dust and hence reaches higher equilibrium temperatures. It mainly
cools through FIR/submm line radiation via atomic fine structure lines
and molecular rotation lines.

\begin{figure}
\centering
\includegraphics[height=0.54\textwidth]{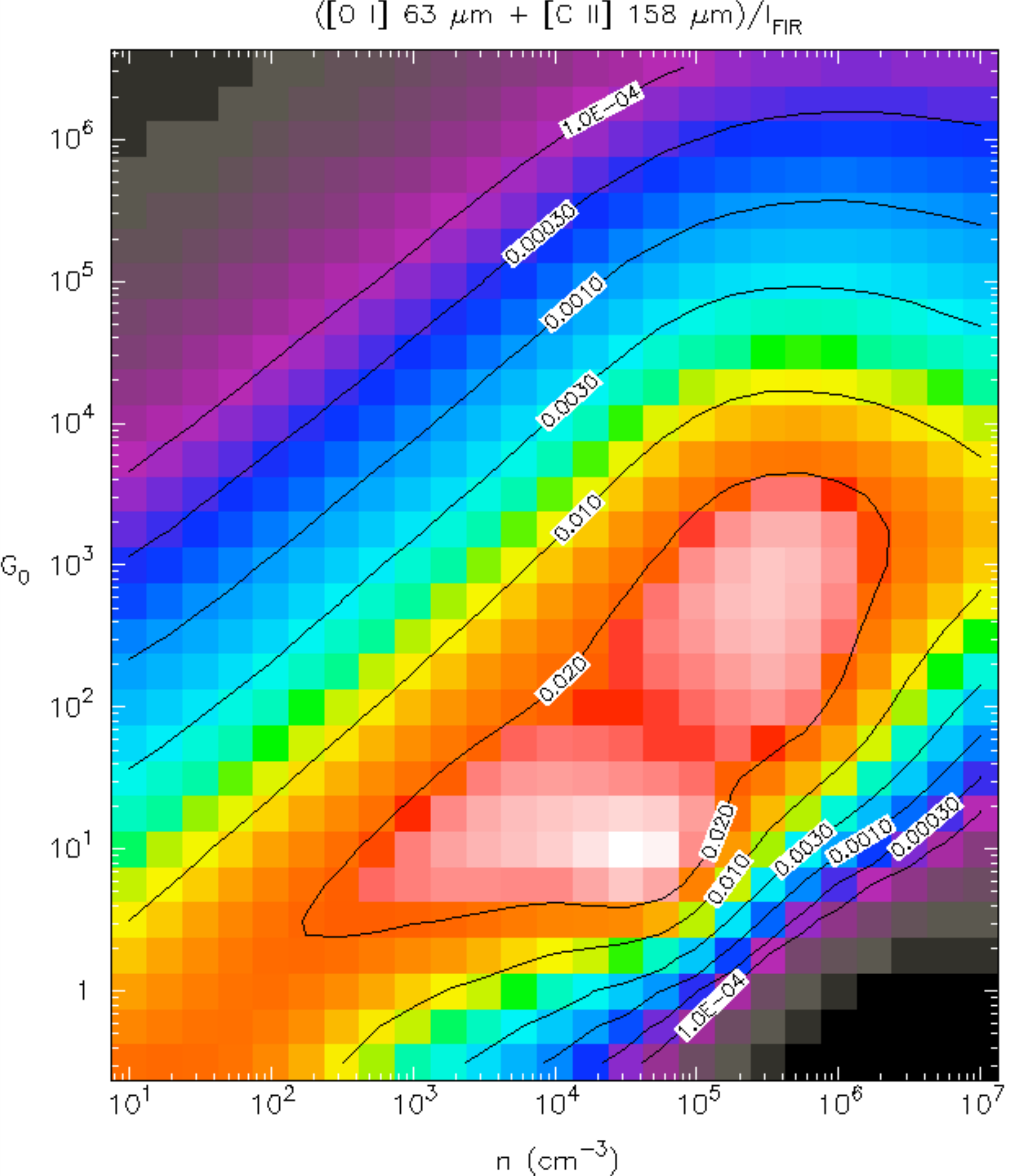}
\includegraphics[height=0.54\textwidth]{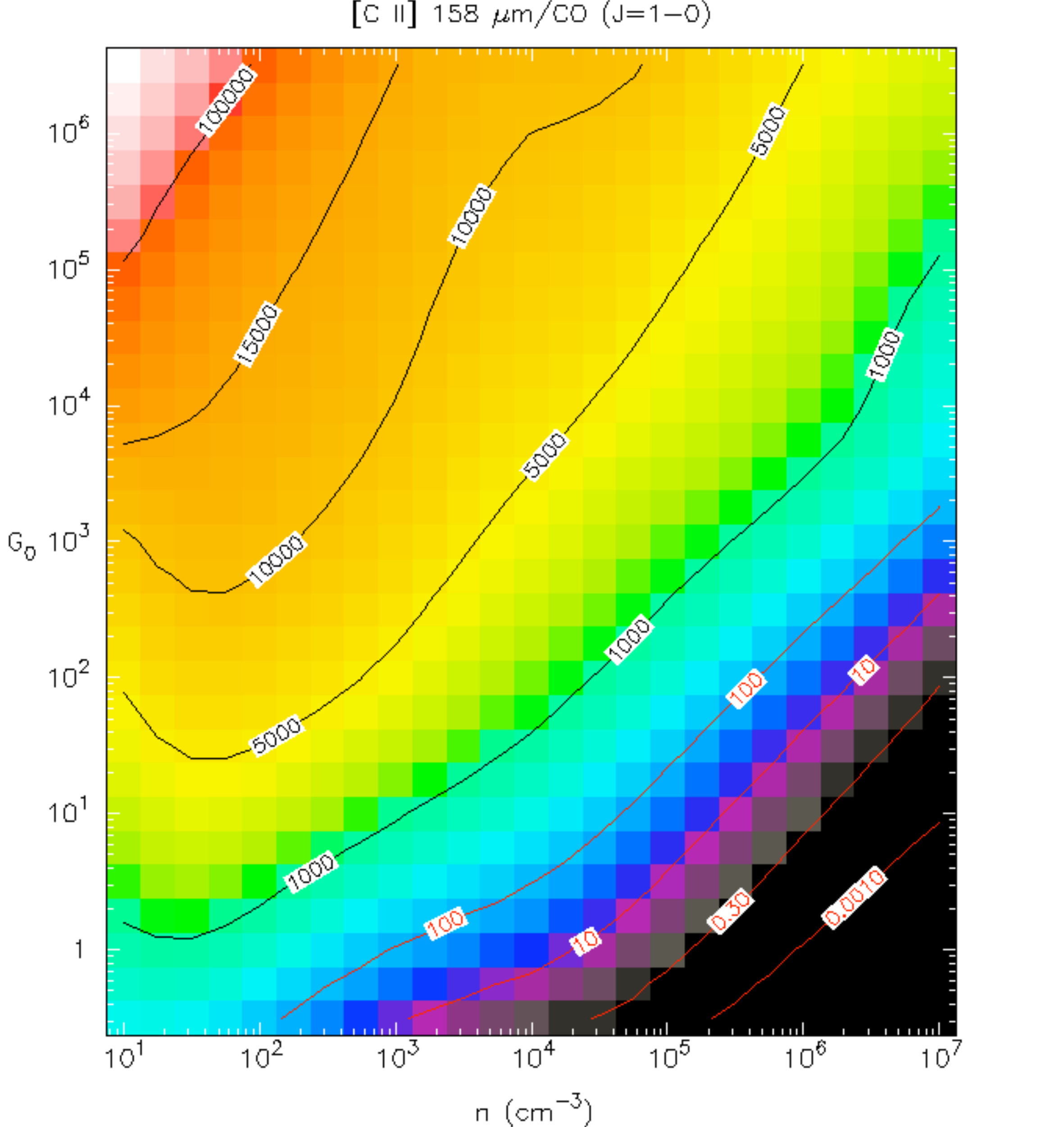}
\caption{Theoretical calculation
  $L_{\text{[C{\sc{ii}}]+[O{\sc{i}}]}}/L_{\text{FIR}}$ and
  $L_{\text{[C{\sc{ii}}]}}/L_{\text{CO(1-0)}}$ of PDRs as a function of
  the hydrogen number density and the strength of FUV radiation field
  falling onto the gas. Figures taken from the PDR Toolbox
  (http://dustem.astro.umd.edu/pdrt/).}
\label{PDRT.pdf}
\end{figure}

The first theoretical studies of PDR physics and chemistry in the
1970s concentrated on ``obvious'' PDR regions such as diffuse clouds
or translucent clouds \citep{1974ApJ...193...73G,
  1976ApJ...203..132B}. Nowadays, PDR studies include a much wider
field, including the pervasive warm neutral medium, giant molecular
clouds and the ISM in the cores of starburst and/or active galaxies. A
plethora of PDR theoretical models have emerged during the past two
decades \citep[e.g.][]{1985ApJ...291..722T, 1986ApJS...62..109V,
  1989ApJ...338..197S, 1996ApJ...468..269D, 1999ApJ...527..795K}. PDR
models typically are advanced computer codes accounting for a growing
number of physical and chemical effects with increasing accuracy. PDR
codes simultaneously solve for the relative abundances and level
populations of different species, radiative transfer and thermal
balance equations in a given geometry. Important ingredients are the
grain and PAH properties, the choice of the species in the gas
mixtures, the chemical reactions, the heating and cooling mechanisms
and the geometry. State-of-the-art codes are now routinely used to
make detailed model predictions for observables such as fine-structure
line intensities and intensity ratios as a function of the main
physical parameters such as the strength of the FUV radiation field
and the hydrogen number density (Figure~{\ref{PDRT.pdf}}).

From such studies, we can determine the most important cooling lines
and create powerful gas diagnostics. While the theoretical PDR models
nowadays make detailed predictions of the strengths of all these
lines, it is important to be aware of the limitations of these
models. Several state-of-the-art numerical PDR codes were recently
compared and benchmarked by \citet{2007A&A...467..187R}. One of the
main results was that a significant spread remains between the
computed observables, such that caution is needed when comparing
astronomical data with PDR model predictions.

\section{The far-infrared properties of galaxies}

\subsection{FIR spectroscopy of nearby galaxies}

%\begin{figure}
%\centering
%\includegraphics[height=0.50\textwidth]{Fischer.pdf}
%\caption{.}
%\label{Fischer.pdf}
%\end{figure}

\begin{figure}
\centering
\includegraphics[height=0.45\textwidth]{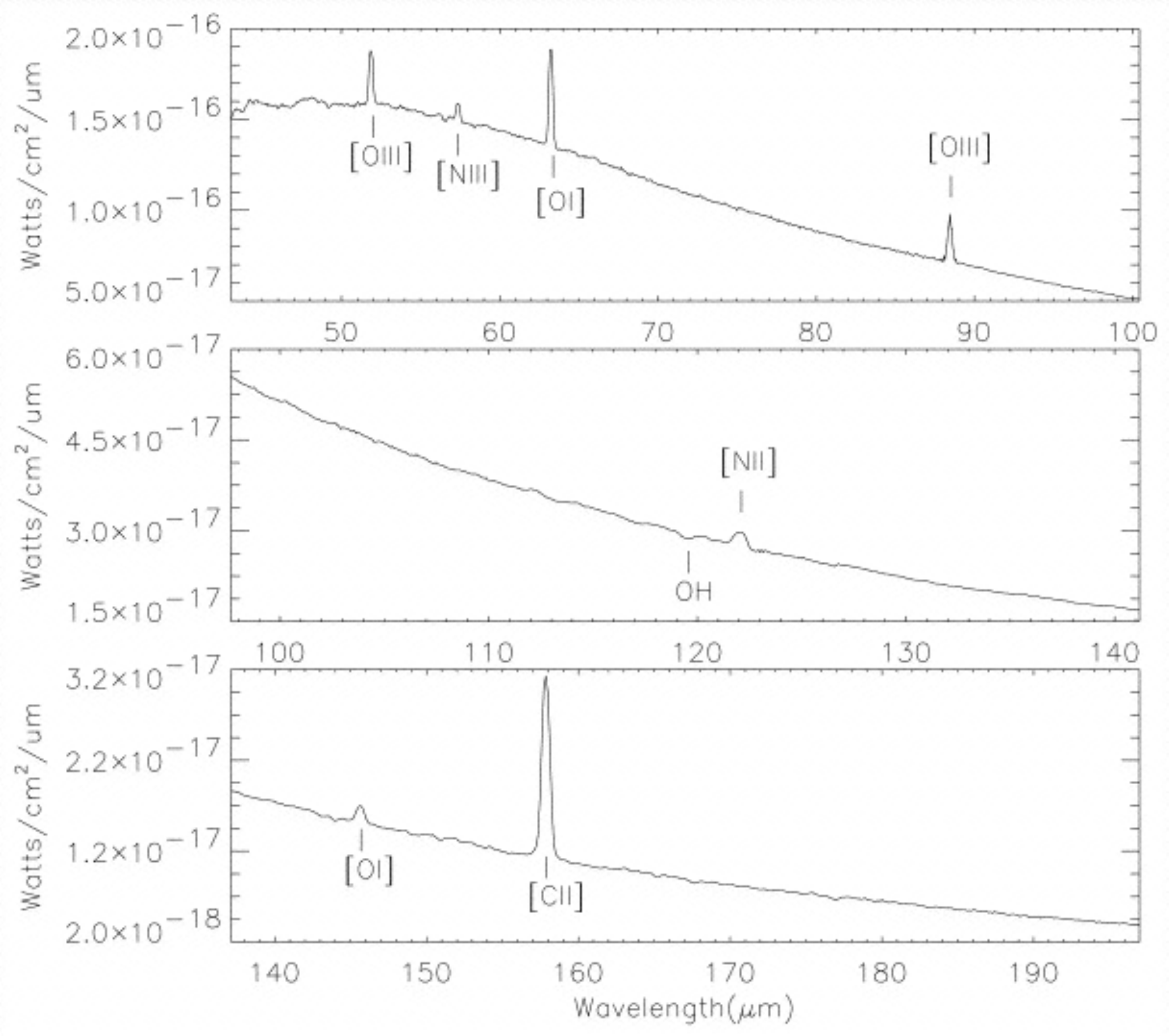}
\includegraphics[height=0.45\textwidth]{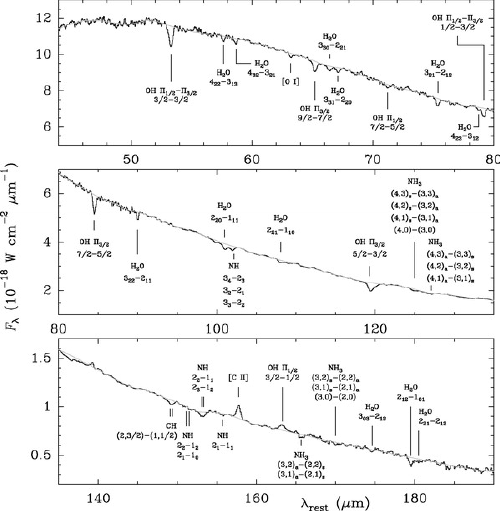}
\caption{ISO LWS spectra of M82 (left) and Arp\,220 (right). In the
  starburst galaxy M82, fine-structure emission lines corresponding to
  both neutral and ionized gas are present
  \citep{1999ApJ...511..721C}. The spectrum of the optically thick
  ULIRG Arp\,220 contains many signatures of molecular transitions, as
  well as [C{\sc{ii}}] in emission and [O{\sc{i}}] in absorption
  \citep{2004ApJ...613..247G}.}
\label{FIRspectra.pdf}
\end{figure}

The first FIR spectroscopic observations were done with KAO, COBE and
balloons in the 1980 and early 1990s. A major step forward was the
appearance of ISO with its LWS spectrograph, which made studies of
large samples of galaxies of different types possible
\citep[e.g.][]{1997ApJ...491L..27M, 2001ApJ...561..766M,
  1999MNRAS.310..317L, 2001A&A...375..566N}.

In agreement with model predictions, the fine-structure line of singly
ionized carbon [C{\sc{ii}}] at 158 $\mu$m, is the single most
important cooling line of the neutral ISM of normal starforming
galaxies (Figure~{\ref{FIRspectra.pdf}}a). Galaxies typically have
$L_{\text{[C{\sc{ii}}]}}/L_{\text{FIR}}$ in the range 0.1 to 1\%,
consistent with PDR models (Figure~{\ref{PDRT.pdf}}a). With an
ionization energy of 11.2~eV, C{\sc{ii}} exists mainly in the neutral
ISM, but a fraction of the [C{\sc{ii}}] emission can originate from
diffuse ionized regions \citep{1993ApJ...407..579M,
  1994ApJ...434..587B, 2002A&A...382..600M}. Other important
fine-structure lines include the [O{\sc{i}}] lines at 63 and 145
$\mu$m and the [C{\sc{i}}] lines at 370 and 609~$\mu$m. The
[O{\sc{i}}] line becomes a more efficient cooling line than
[C{\sc{ii}}] in high-density regions. The ratios of all these
different lines can be used to determine the temperature, density,
filling factor and FUV radiation field strength.

The FIR spectra of starforming galaxies also contain a number of
atomic fine-structure lines that trace ionized gas. The [N{\sc{ii}}]
lines at 122 and 205~$\mu$m are good tracers of diffuse low-density
ionized gas, whereas the [N{\sc{iii}}] line at 57~$\mu$m and the
[O{\sc{iii}}] lines at 52 and 88~$\mu$m trace the denser ionized
gas. Ratios of these lines can be used to derive electron density and
the hardness of the UV radiation field.

Finally, the FIR/submm region contains many transitions of molecular
lines, including CO, OH, H$_2$O and CH
(Figure~{\ref{FIRspectra.pdf}}b). Molecular line emission originates
both in PDRs and in denser molecular cores. The luminosity in the
CO(1-0) line is typically three or more orders of magnitude lower than
$L_{\text{[C{\sc{ii}}]}}$ for normal starforming galaxies
(Figure~{\ref{PDRT.pdf}}b).

%\begin{figure}
%\centering
%\includegraphics[height=0.50\textwidth]{ratio-fir.pdf}
%\caption{Figure from \citet{2009A&A...497..351C}.}
%\label{ratio-fir.pdf}
%\end{figure}

\subsection{Unsolved issues concerning [C{\sc{ii}}] emission}

Somewhat surprisingly, galaxies with higher $L_{\text{FIR}}$ and
warmer infrared colours have smaller
$L_{\text{[C{\sc{ii}}]}}/L_{\text{FIR}}$ ratios. In particular,
luminous and ultraluminous infrared galaxies show a clear deficiency
in [C{\sc{ii}}] \citep{1998ApJ...504L..11L,
  2001ApJ...561..766M}. Several explanations have been proposed to
explain this trend. The most common explanation is an increased grain
charge in galaxies with high levels of FUV radiation, which creates a
higher potential barrier to photoelectric ejection and hence reduces
the efficiency of the photoelectric heating
\citep{1998ApJ...504L..11L}. However, this issue is far from settled:
other teams propose other explanations, including an increase in the
collisional de-excitation in the [C{\sc{ii}}] transition at high
density \citep{2001A&A...375..566N} or non-PDR contributions to the
FIR continuum \citep{2003ApJ...594..758L, 2009A&A...497..351C}.

Another tentalizing discovery by ISO LWS is that low-metallicity
galaxies, including both dwarf irregulars and blue low-mass spiral
galaxies, often exhibit unusually intense [C{\sc{ii}}] emission as
compared with the 2.6~mm CO(1-0) emission \citep{1994ApJ...430L..37M,
  1995ApJ...454..293P, 1997ApJ...483..200M, 1997AJ....114..138S,
  2001ASPC..231..236M}. This result may be due to low abundances of
dust and heavy elements: in such gas, FUV radiation penetrates more
deeply into a molecular cloud, causing a larger C{\sc{ii}} region
relative to the CO core. Alternatively, radiation from diffuse
H{\sc{i}} clouds may dominate the [C{\sc{ii}}] emission from these
galaxies. These observations are a strong indication that CO(1-0)
measurements are not always a faithful proxy for the molecular gas
content.

\subsection{Galaxies at high redshift}

In the past decade, submm spectroscopy of galaxies at cosmological
distances has seen a spectacular development
\citep{2005ARA&A..43..677S, 2007RPPh...70.1099O}.  Most of this work
was done by the IRAM Plateau de Bure and 30m telescopes, thanks to
both improved detector technology and broader spectral bandwidth which
have made the detection of faint extragalactic emission lines with
widths of several hundred up to a thousand km/s possible. Most
molecular line studies concentrate on the various transitions of the
CO lines, which are our best tracer of molecular hydrogen, and
therefore an ideal tracer of the gas reservoirs needed to sustain the
prodigious star formation rates implied by their strong far-IR dust
continuum emission. The interpretation of CO emission requires
observations at various transitions to populate the ``CO ladders'' and
disentangle the temperature and density effects
\citep{2007A&A...467..955W} and still limits to densities up to
10$^5$~cm$^{-3}$. Other molecular tracers such as HCN have since been
detected in the brightest high-redshift objects.

\begin{figure}
\includegraphics[height=0.50\textwidth]{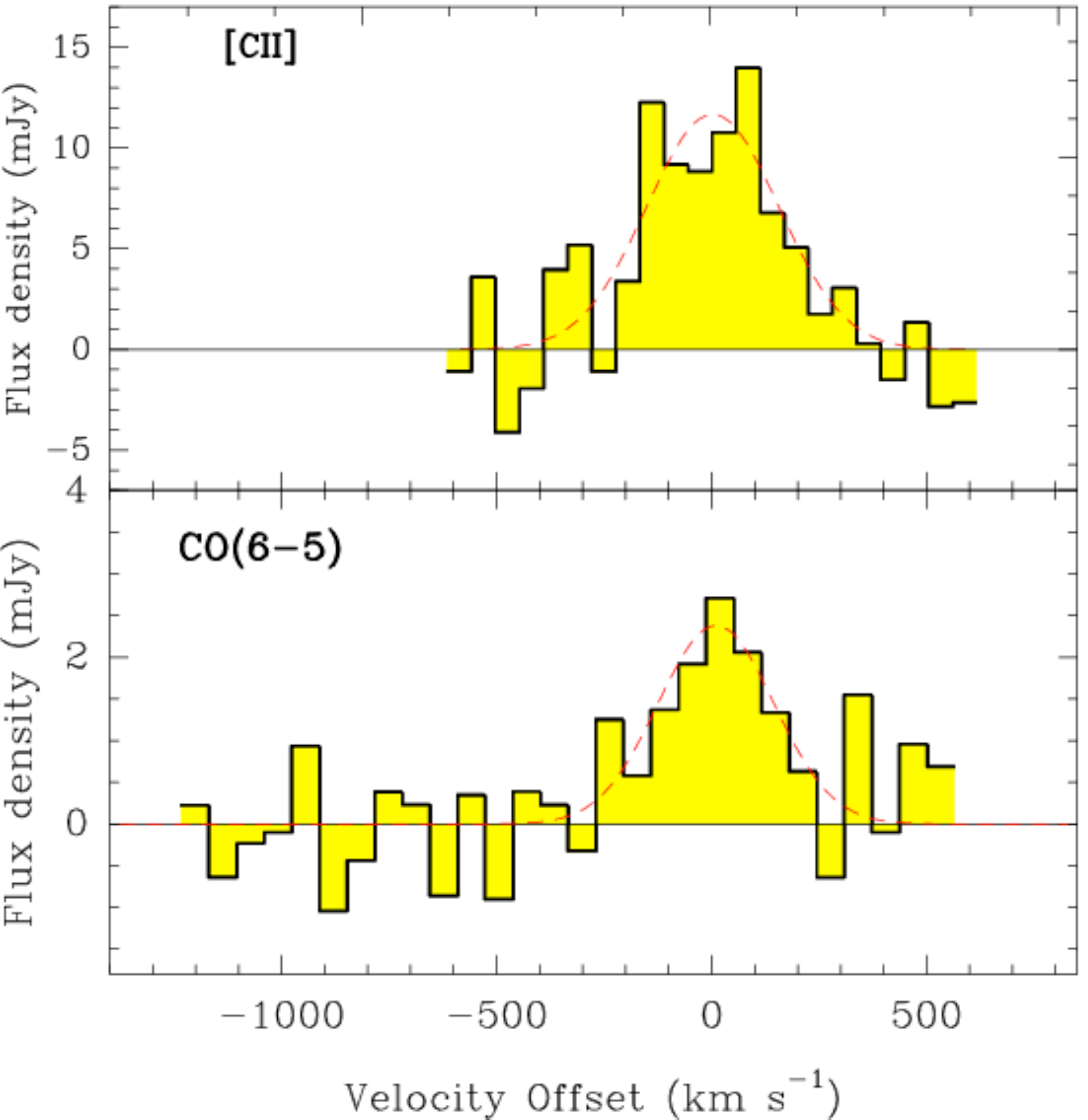}
\includegraphics[height=0.50\textwidth]{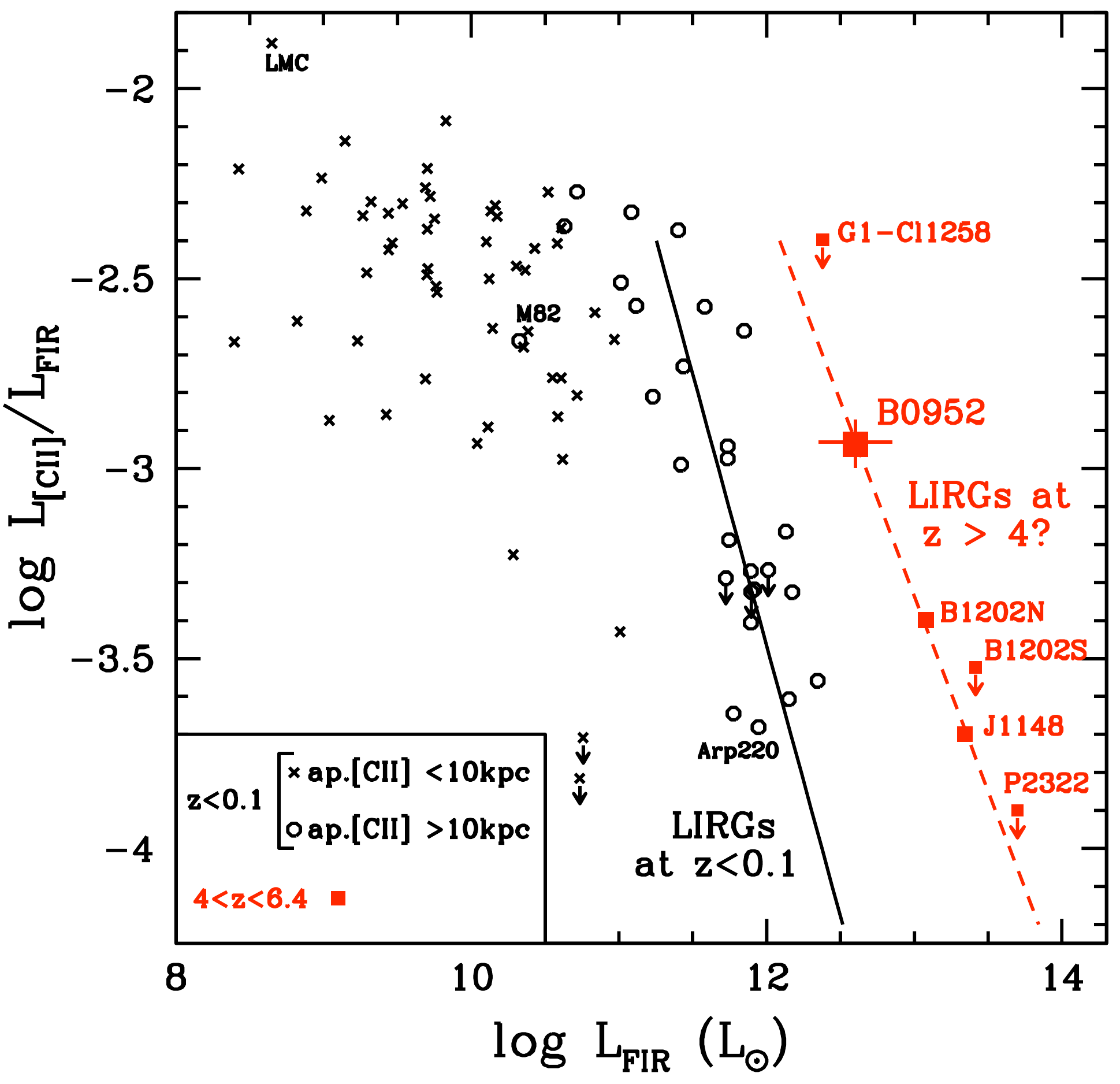}
\caption{Left: the detection of [C{\sc{ii}}] and CO(6-5) emission in a
  quasar at $z=6.42$ using the IRAM 30m telescope
  \citep{2005A&A...440L..51M}. Right: the
  $L_{[{\sc{ii}}]}/L_{\text{FIR}}$ ratio versus $L_{\text{FIR}}$ for
  normal and starburst local galaxies (black crosses and circles) and
  three high-redshift sources (red squares) as in
  \citet{2009arXiv0904.3793M}.}
\label{CIIhighz.pdf}
\end{figure}

The study of high redshift objects also allows us to observe the
[C{\sc{ii}}] line from ground-based telescopes. So far, [C{\sc{ii}}]
emission has been detected in one quasar at $z=6.42$
\citep{2005A&A...440L..51M, 2009Natur.457..699W} at mm wavelengths and
in two systems at $4<z<5$ at submm wavelengths
\citep{2006ApJ...645L..97I, 2009arXiv0904.3793M}
(Figure~{\ref{CIIhighz.pdf}}a). Particularly interesting is the recent
detection of [C{\sc{ii}}] emission in a lensed system at $z=4.43$,
detected by APEX, which has a superb transmission at submm wavelengths
thanks to its location on the Chajnantor plane (the future ALMA
site). This observation suggests that the [C{\sc{ii}}] line is
brighter than intially assumed, and may be offset from the local
$L_{\text{[C{\sc{ii}}]}}/L_{\text{FIR}}$ relation
(Figure~{\ref{CIIhighz.pdf}}b). This finding remains to be confirmed
by future observations, and the physical interpretation is not clear.

\section{The promise of Herschel}

\subsection{The Herschel Space Observatory}

The Herschel Space Observatory, the fourth cornerstone mission in the
ESA science programme, was successfully launched on 14 May 2009. With
a 3.5 m Cassegrain telescope it is the largest space telescope ever
launched. It will perform photometry and spectroscopy in the FIR/submm
wavelength range between approximately 55 and 672 $\mu$m, bridging the
gap between earlier infrared space missions and ground-based
facilities. Herschel is the only space facility dedicated to the
FIR/submm spectral regime. Its vantage point in space provides several
decisive advantages, including a low and stable background and full
access to this part of the spectrum.

Herschel has three scientific instruments on board.  PACS
\citep{2009EAS....34...43P} is a camera and low to medium resolution
spectrometer for wavelengths in the range between 55 and 210 $\mu$m
($R\sim1500$).
%It employs four detector arrays, two bolometer arrays and two Ge:Ga photoconductor arrays. The bolometer arrays are dedicated for wideband photometry, while the photoconductor arrays are to be employed exclusively for spectroscopy with a resolution of a few thousand. PACS can be operated either as an imaging photometer, or as an integral field line spectrometer. PACS is built by a consortium led by MPE in Garching 
SPIRE \citep{2009EAS....34...33G} is a camera and low to medium
resolution spectrometer complementing PACS for wavelengths in the
range 194--672 $\mu$m ($R\lesssim1000$ at 250~$\mu$m). It comprises an
imaging photometer and a Fourier Transform Spectrometer (FTS), both of
which use bolometer detector arrays.
%There are a total of five arrays, three dedicated for photometry and two for spectroscopy. SPIRE is built by a consortium led by Cardiff University. 
Finally, HIFI \citep{2009EAS....34....3D} is a very high resolution
heterodyne spectrometer covering the 490-1250 GHz and 1410-1910 GHz
bands ($R\sim3\times10^6$).
%It utilises low noise detection using superconductor-insulator-superconductor (SIS) and hot electron bolometer (HEB) mixers, together with acousto-optical and autocorrelation spectrometers. HIFI is not an imaging instrument, it observes a single pixel on the sky at a time. It is built by a consortium led by SRON in Groningen. 
The three instruments are nationally funded (by the ESA member states
with contributions from the USA, Canada and Poland).

\subsection{Extragalactic FIR spectroscopy with Herschel}

The operational lifetime of Herschel is foreseen to be 3 years; about
one third of the time is dedicated for guaranteed time observations,
two thirds is available for the community through open time. For both
guaranteed and open time, a large fraction of the time is used for key
programmes (with more than 100 hours of observing time). There are
five guaranteed time key programs planned on nearby galaxies (one of
them only uses submm imaging, the others use a combination of
FIR/submm imaging and spectroscopy). The general topic of these
proposals is the evolution of the different phases (the stellar, dusty
and gaseous phases) within galaxies. The FIR/submm wavelengths probed
by Herschel are absolutely crucial for understanding the physical
processes and properties of the interstellar medium, the interplay
between star formation and the interstellar medium in galaxies, and
how they may depend on the wider galaxian environment.

\begin{figure}
\centering
\includegraphics[height=0.36\textwidth]{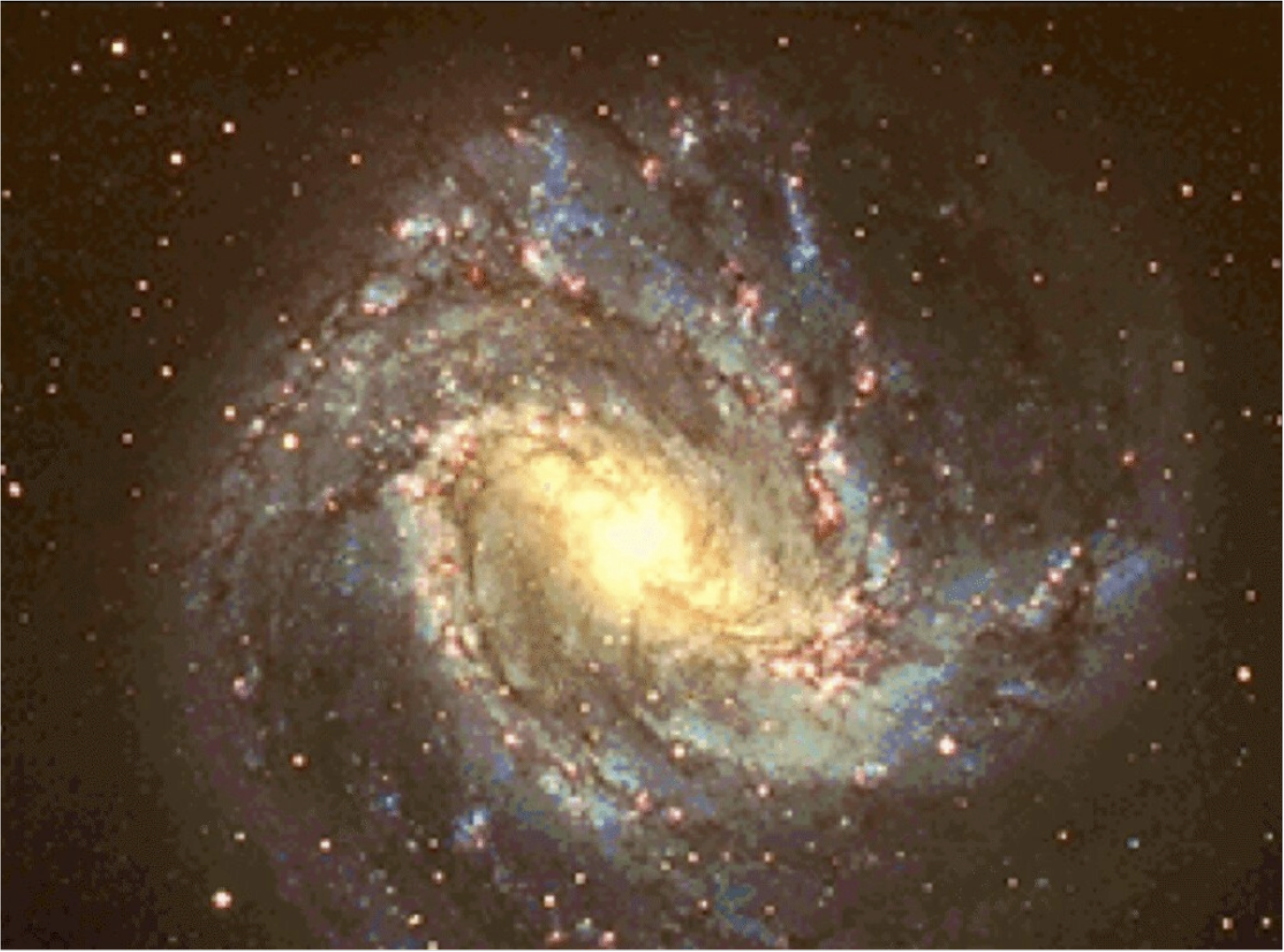}
\includegraphics[height=0.36\textwidth]{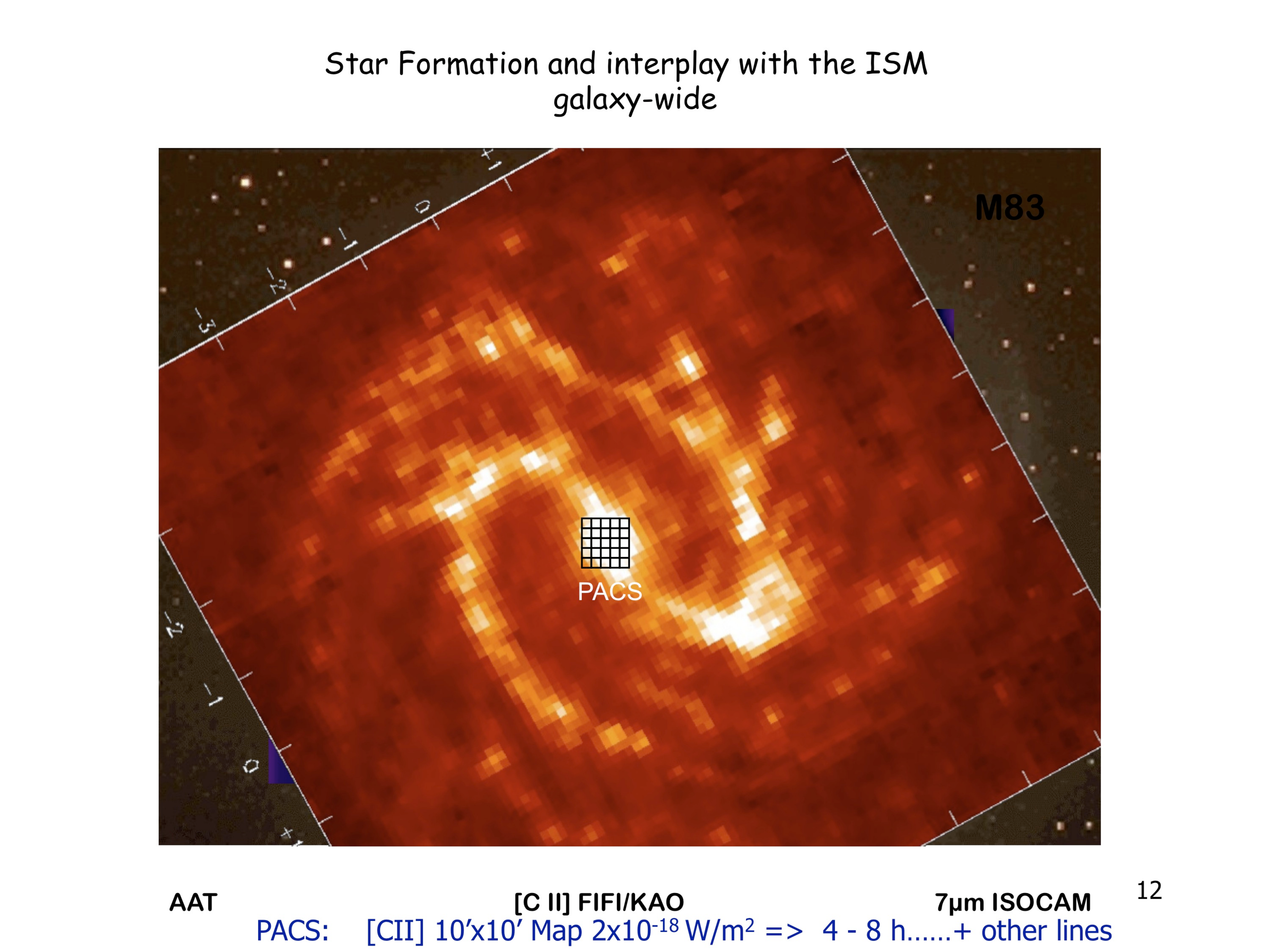}
\caption{M83 as an example of a target of the Herschel VNG key
  program. The left panel shows an optical image, the right panel
  shows a 7~$\mu$m ISOCAM image \citep{2005A&A...441..491V}. The FOV
  and pixel size of the PACS spectroscopy array are also indicated in
  the right panel. Thanks to the superb resolution, we can investigate
  the ISM independently in the nucleus, the arm and the interarm
  regions.}
\label{M83.pdf}
\end{figure}

The Very Nearby Galaxies (VNG) key program by the SPIRE consortium
will use SPIRE and PACS to measure the emission spectrum from dust as
well as important cooling lines from the gaseous ISM in sample of 13
very nearby, prototypical galaxies (M51, M81, NGC\,2403, NGC\,891,
M83, M82, Arp\,220, NGC\,4038/39, NGC\,1068, NGC\,4151, Cen\,A,
NGC\,4125, and NGC\,205). These galaxies have been chosen to probe as
wide a region in galaxy parameter space as possible while maximizing
the achievable spatial resolution and are already well-studied from
X-ray and optical through to radio wavelengths. Since the galaxies are
well-resolved by the PACS and SPIRE beams, the program will allow to
study both variations inside individual galaxies and compare global
properties between different galaxies.

Another SPIRE consortium key program will probe the ISM of low
metallicity environments. As ISO LWS observations suggest, metallicity
can have a profound influence on the ISM structure, on the dust
properties, on the radiation field and on the star formation
activity. The low-metallicity program will use PACS and SPIRE to map
the dust and gas in a 51 dwarf galaxies, sampling a broad metallicity
range of 1/50 to 1/3 $Z_\odot$. These data, in conjunction with other
ancillary data, will be used to construct the emission spectrum of the
dust plus that of the gas in the most important cooling lines. Since
low-metallicity dwarf galaxies are the best Local Universe analogs to
high-redshift protogalaxies, the interpretation of this data will open
the door to comprehending primordial ISM conditions and star formation
in the young universe.

The SHINING program is a PACS consortium key program aiming at a FIR
spectroscopic and photometric survey of infrared bright galaxies at
$0<z<1$. The aim is to obtain a comprehensive view of the physical
processes at work in the ISM of local galaxies ranging from objects
with moderately enhanced star formation to the most dense, energetic,
and obscured environments in ultra-luminous infrared galaxies (ULIRGs)
and around AGN.  The objects cover a wide parameter range in
luminosity, activity level, and metal enrichment, and will be
complemented by a few objects at intermediate redshifts, i.e. at a
more active epoch of star formation. The interpretation of the data
will be based on a combination of PDR modelling of the neutral and
molecular regions and photoionization modelling of the H{\sc{ii}}
regions.

Finally, the HIFI consortium key program called HEXGAL aims at a high
spectral resolution submm study of the nuclei of a sample of nearby
galaxies (including our own Milky Way). Apart from the key FIR
fine-structure lines that will be mapped with PACS, the HEXGAL program
focuses on the bright fine-structure lines of atomic carbon, a unique
set of water lines, and the high-excitation CO transitions. The
multi-line data will be combined with numerical radiative transfer and
chemical network models quantitatively constrain the various phases of
the ISM.

\section{Conclusion}

The FIR/submm region is a fascinating wavelength domain that holds the
key to study both the neutral, ionized and molecular phases of the
ISM. Limited spectroscopic studies so far, mainly based on ISO LWS
observations, have already demonstrated the potential of this
wavelength region. There is no doubt that the new Herschel mission
will revolutionize the study of the ISM of nearby galaxies in terms of
sensitivity, spatial and spectral resolution in the coming few
years. On a slightly further baseline, ALMA will do a similar job at
higher redshift. It will allow us to study the chemisty of the
primordial gas in galaxies, which is a strong side constraint on the
cosmic evolution of star formation and chemical evolution of galaxies.



\end{document}